\documentclass[twocolumn,aps,superscriptaddress]{revtex4-1}
\usepackage{amsmath}
\usepackage{bm}
\usepackage{graphicx}
\usepackage{dcolumn}
\usepackage{epstopdf}
\usepackage{epsfig}
\usepackage{color}
\usepackage{float}

\usepackage[colorinlistoftodos,prependcaption]{todonotes}

\begin{document}
\title{Optimization of single crystal growth of candidate quantum spin-ice \bm{${\rm Pr_2Hf_2O_7}$} by optical floating-zone method }
\author{V.\ K.\ Anand}
\altaffiliation{vivekkranand@gmail.com}
\author{\mbox{A.\ T.\ M.\ N.\ Islam}}
\affiliation{\mbox{Helmholtz-Zentrum Berlin f\"{u}r Materialien und Energie GmbH, Hahn-Meitner Platz 1, D-14109 Berlin, Germany}}
\author{A.\ Samartzis}
\author{J.\ Xu}
\affiliation{\mbox{Helmholtz-Zentrum Berlin f\"{u}r Materialien und Energie GmbH, Hahn-Meitner Platz 1, D-14109 Berlin, Germany}}
\affiliation{\mbox{Institut f\"{u}r Festk\"{o}rperphysik, Technische Universit\"{a}t Berlin, Hardenbergstra{\ss}e 36, D-10623 Berlin, Germany}}
\author{N.\ Casati}
\affiliation{\mbox{Paul Scherrer Institute, 5232 Villigen PSI, Switzerland}}
\author{B.\ Lake}
\altaffiliation{bella.lake@helmholtz-berlin.de}
\affiliation{\mbox{Helmholtz-Zentrum Berlin f\"{u}r Materialien und Energie GmbH, Hahn-Meitner Platz 1, D-14109 Berlin, Germany}}
\affiliation{\mbox{Institut f\"{u}r Festk\"{o}rperphysik, Technische Universit\"{a}t Berlin, Hardenbergstra{\ss}e 36, D-10623 Berlin, Germany}}

\date{\today}

\begin{abstract}
  We report the optimized conditions for growing the high quality single crystals of candidate quantum spin-ice Pr$_2$Hf$_2$O$_7$ using the optical floating-zone method. Large single crystals of Pr$_2$Hf$_2$O$_7$ have been grown under different growth conditions using a four-mirror type optical floating-zone furnace and their microscopic structural differences have been probed by high-resolution synchrotron x-ray diffraction (SXRD). The SXRD data reveal that the crystals grown under flowing argon ($\sim 2$~L/h) atmosphere with slightly off-stoichiometric (optimized) starting composition yields the highest quality crystals. The magnetic susceptibility, isothermal magnetization and heat capacity data of optimally grown crystals are presented.
\end{abstract}

\maketitle

\section{\label{Intro} INTRODUCTION}

The 227 rare earth pyrochlores have been a topic of interest owing to their exotic properties, for example Dy$_2$Ti$_2$O$_7$ and Ho$_2$Ti$_2$O$_7$ present spin-ice behavior for which magnetic monopoles are believed to be the fundamental excitations \cite{Gardner2010, Castelnovo2012, Gingras2014, Ramirez1999, Siddharthan1999,Hertog2000,Bramwell2001,Castelnovo2008}. In these materials rare earth ions form a three dimensional network of corner-sharing tetrahedra where ferromagnetic interaction becomes frustrated under the action of the crystal electric field (CEF) that forces the spins to point along the local $\langle 111 \rangle $ direction. The CEF induced Ising anisotropic ground state of the classical spin-ice materials consists of `two-in/two-out' spin configuration referred as `ice rule' \cite{Harris1997}. Depending upon the relative strengths of CEF anisotropy and the nearest neighbor dipolar and exchange interactions these materials exhibit a wide range of magnetic ground states. For example, spin-ice behavior when ferromagnetic dipolar interaction dominates over the antiferromagnetic exchange interaction as in Dy$_2$Ti$_2$O$_7$ and Ho$_2$Ti$_2$O$_7$ \cite{Ramirez1999, Siddharthan1999, Hertog2000, Bramwell2001}, and all-in/all-out antiferromagnetic order when antiferromagnetic exchange interaction dominates such as in Nd$_2$Zr$_2$O$_7$ \cite{Lhotel2015,Xu2015} and Nd$_2$Sn$_2$O$_7$ \cite{Bertin2015}. 

In our efforts to extend work on rare earth pyrochlores, we recently started working on hafnate pyrochlores  \cite{Anand2015,Anand2017,Anand2016} which remained poorly investigated.  Our investigations revealed a long-range antiferromagnetic ordering in Nd$_2$Hf$_2$O$_7$ ($T_{\rm N }\approx 0.55$~K) with an all-in/all-out arrangement of Nd$^{3+}$ moments determined by the neutron diffraction study \cite{Anand2015}. The neutron diffraction data showed a strongly reduced ordered moment of 0.62(1)~$\mu_{\rm B}$/Nd at 0.1~K, much lower than the expected 2.5~$\mu_{\rm B}$/Nd for the Ising anisotropic ground state of Nd$_2$Hf$_2$O$_7$. Persistent dynamic spin fluctuations deep inside the ordered state were found in the muon spin relaxation study which seems to be responsible for the strongly reduced ordered moment \cite{Anand2017}. The inelastic neutron scattering study revealed the Kramers doublet ground state of Nd$_2$Hf$_2$O$_7$ to be of dipolar-octupolar type  \cite{Anand2017}.

Very recently the hafnate pyrochlore Pr$_2$Hf$_2$O$_7$ was identified to be a potential candidate quantum spin-ice system \cite{Anand2016,Sibille2016}. Pr$_2$Hf$_2$O$_7$ was found to show no long-range magnetic ordering down to 90~mK, however the frequency dependent behavior of ac magnetic susceptibility clearly reflects slow spin dynamics and spin-freezing behavior \cite{Anand2016,Sibille2016}. While the magnetic ground state of Pr$_2$Hf$_2$O$_7$ is Ising anisotropic in nature, the presence of a weak non-Ising contribution is also inferred from the isothermal magnetization data \cite{Anand2016} thus favoring  dynamic spin-ice behavior in Pr$_2$Hf$_2$O$_7$. Further, Pr$_2$Hf$_2$O$_7$ has a non-Kramers doublet ground state which is considered favorable for the dynamic spin-ice behavior \cite{Lee2012}.   

In order to probe and understand the intrinsic properties of Pr$_2$Hf$_2$O$_7$ in more detail it is highly desired to have a high quality single crystal of this compound. Because of the higher melting point of hafnate pyrochlores (compared to titanate pyrochlores) the synthesis of large single crystals of Pr$_2$Hf$_2$O$_7$ using optical floating-zone method turns out to be somewhat tricky. While one can grow titanate pyrochlores in air at ambient pressure, the same recipe does not work for Pr$_2$Hf$_2$O$_7$ which tends to decompose to Pr$_6$O$_{11}$ and HfO$_2$. As such one needs to grow Pr$_2$Hf$_2$O$_7$ under a pressurized/argon atmosphere. Very recently Ciomaga Hatnean {\it et al}. \cite{Hatnean2017} also reported a successful growth of Pr$_2$Hf$_2$O$_7$ single crystal under pressure. We find that the quality of crystals not only depends on the atmospheric conditions but also on the stoichiometry of the starting material. We have optimized the growth conditions to yield a high quality crystal as report here. We used high-resolution synchrotron x-ray diffraction (SXRD) measurements to distinguish the subtle structural differences of crystals grown under different growth conditions. We also present the magnetic susceptibility $\chi(T)$, isothermal magnetization $M(H)$ and heat capacity $C_{\rm p}(T)$ data of single crystal grown under optimized condition. The Ising anisotropy is confirmed, and the bulk properties are found to be consistent with those of polycrystalline Pr$_2$Hf$_2$O$_7$ \cite{Anand2016}.

\section{\label{ExpDetails} EXPERIMENTAL DETAILS}

\begin{figure}
\includegraphics[width=\columnwidth, keepaspectratio]{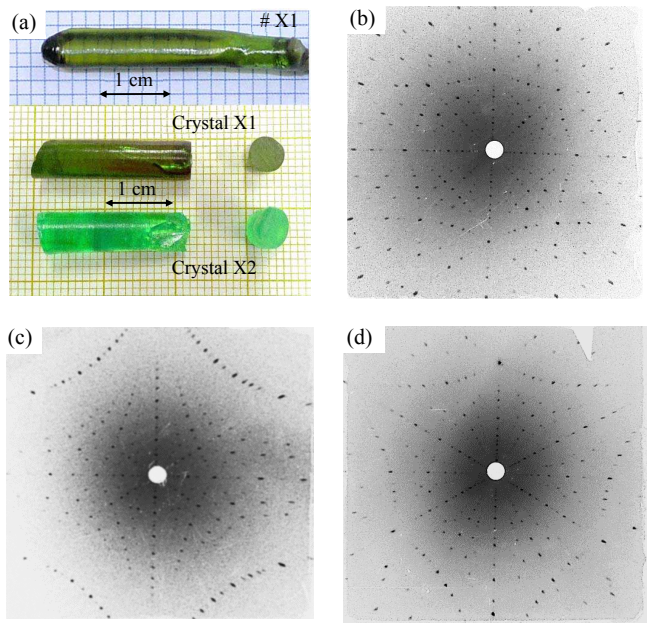}
\caption {(a) Representative single crystal of Pr$_2$Hf$_2$O$_7$ (\# X1) grown by the floating zone method under Ar atmosphere at a rate of 5~mm/h. Also shown is a piece of crystal \# X2, showing the dependence of crystal color on the stoichiometry of starting material. (b--d) Representative X-ray Laue backscattering photographs along the directions [100], [110] and [111]  of an oriented piece of crystal X2.} 
\label{fig:Laue}
\end{figure}

Single crystal growth of Pr$_2$Hf$_2$O$_7$ was carried out by the Floating-zone technique using a four-mirror type high-temperature optical floating zone furnace (Crystal Systems corp.\ FZ-T-12000-X-VII-VPO-PC) equipped with four 3~kW Xenon arc lamps at the Core Lab for Quantum Materials, Helmholtz-Zentrum Berlin (HZB). Crystals were grown under high purity argon atmosphere with a growth rate of 1--5~mm/h. The average power required for the growth was 36 \% of lamp power (four 3~kW lamps). During the growth the seed and feed rods were counter-rotated at 5 and 20 rpm, respectively. The seed and feed rods were prepared from high purity materials: Pr$_6$O$_{11}$ (99.99\%, Alfa Aesar) and HfO$_2$ (99.95\%, Alfa Aesar) mixed and ground thoroughly and fired at 1300~$^\circ$C for 15~h, reground and fired again at 1300~$^\circ$C for 24~h, then powdered and pressed hydrostatically in the form of cylindrical rod. The cylindical rods were then sintered for 60~h at 1500~$^\circ$C. The sintered rods were typically about 5 mm in diameter and 8-10 cm in length. All grindings and heat treatments were done in air. 

For subsequent growths we used single crystals from previous growths as seed crystals to avoid random nucleation and to obtain one large single crystal. Laue diffraction was used to check the quality of the single crystals and orient along specific crystallographic direction. The oriented crystal was then cut into small pieces for magnetic measurements. The crystal quality was also examined by polarized optical microscopy. The chemical composition of crystals were checked by energy dispersive x-ray (EDX) analysis using a scanning electron microscope.

Further, the quality of the single crystals were probed by high-resolution synchrotron x-ray diffraction measurements on crushed crystals performed on the MS-beamline \cite{Willmott2013} at Paul Scherrer Institute (PSI), Switzerland. The crystals were very finely ground and filled into a thin glass capillary of 0.3~mm diameter.  During the measurement the sample was rotated continuously to reduce the effect of preferred orientation and profile shape dependence. Measurements were performed with incident x-rays of 25~keV energy at a temperature of 300~K\@. In order to accurately determine the wavelength of the x-rays (0.4148 {\AA}) and the instrument profile parameters, the standard LaB$_6$ powder (NIST) was also measured under the identical experimental condition. 

The dc magnetic susceptibility $\chi$ versus temperature $T$ and isothermal magnetization $M$ versus magnetic field $H$ measurements were performed using a superconducting quantum interference device vibrating sample magnetometer (SQUID-VSM, Quantum Design, Inc.) and VSM option of a physical properties measurement system (PPMS, Quantum Design, Inc.) at Core Lab for Quantum Materials, HZB. The heat capacity $C_{\rm p}(T)$ was measured by the adiabatic-relaxation technique using the heat capacity option of PPMS at Core Lab for Quantum Materials, HZB.

\section{\label{Growth} Crystal Growth}  

\begin{figure}
\includegraphics[width=\columnwidth, keepaspectratio]{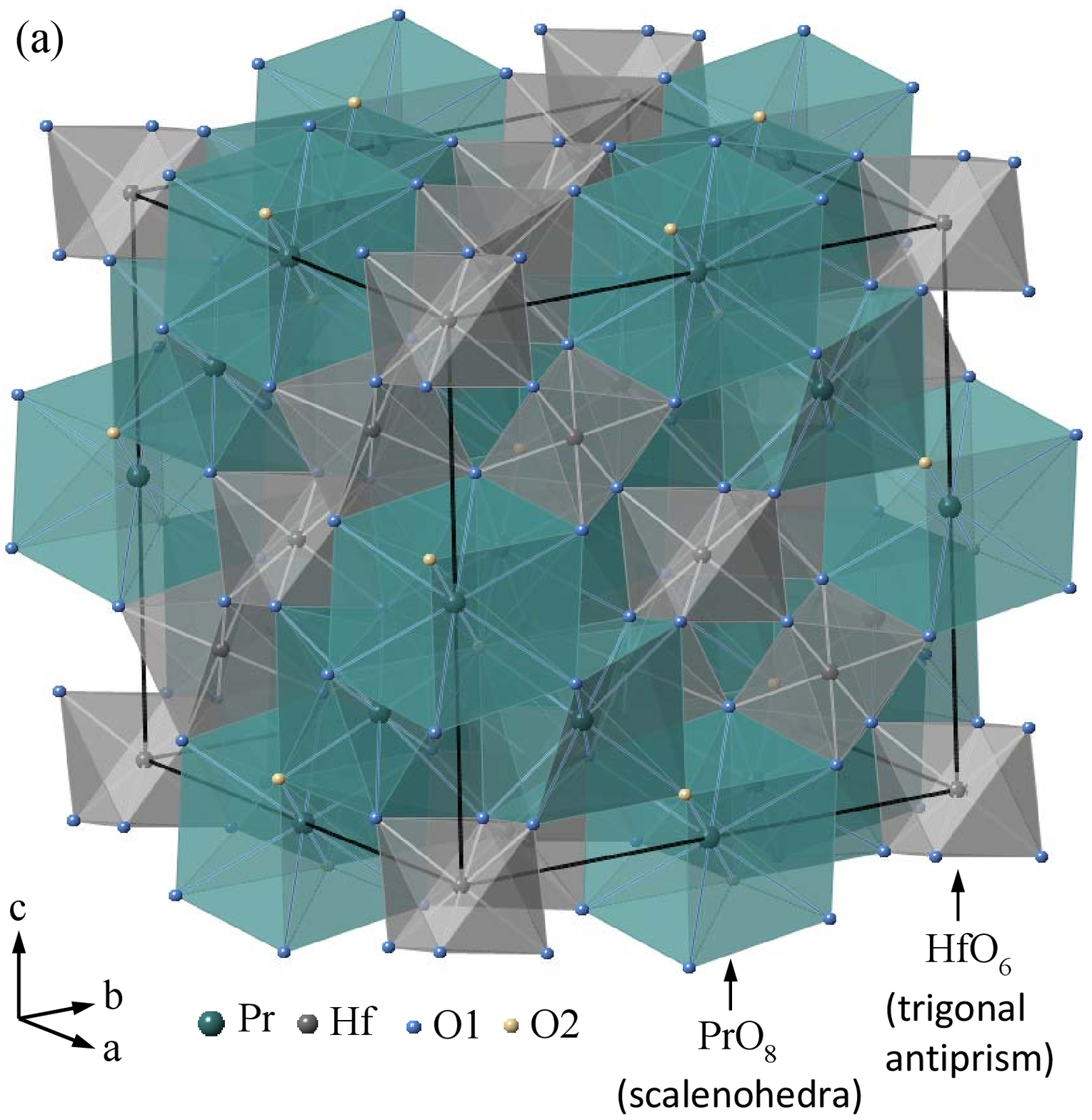}\vspace{0.3cm}
\includegraphics[width=\columnwidth, keepaspectratio]{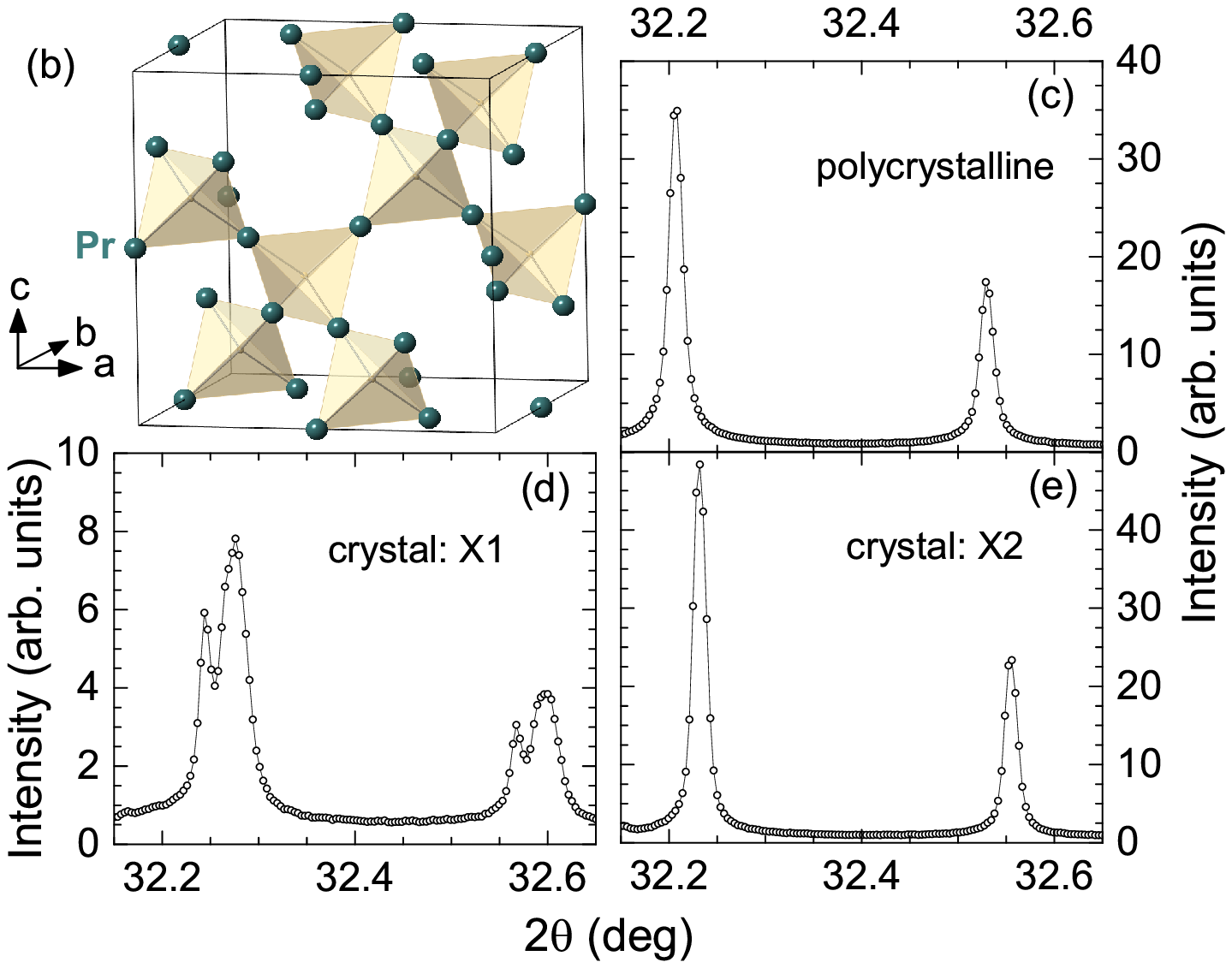}
\caption{(a) Face-centered cubic  (space group $Fd\bar{3}m$) pyrochlore structure of Pr$_2$Hf$_2$O$_7$. Polyhedra around the Pr and Hf atoms show the scalenohedra and trigonal antiprism formed by O atoms. (b) Illustration of corner-sharing tetrahedra formed by Pr atoms. (c--e) A comparison of synchrotron powder x-ray diffraction patterns of crushed single crystals X1 and X2, and polycrystalline sample, shown on an expanded scale (very small $2\theta$ range) to see the differences more clearly.}
\label{fig:Struct}
\end{figure}

The optimization of the crystal growth of Pr$_2$Hf$_2$O$_7$ was done by varying the growth conditions in several subsequent growths. According to the phase diagram of Pr$_2$O$_3$--HfO$_2$ system \cite{Kravchinskaya1978,Shevchenko1984} Pr$_2$Hf$_2$O$_7$ melts incongruently around 2420 $^\circ$C in vacuum. However, the melting behavior in other atmospheric environments remains unclear.  To check the melting behavior of the material in air, in the first growth attempt we used a feed-rod having stoichiometric composition and an ambient air atmosphere. During the growth we observed evaporation of large quantity of black fumes. The black evaporated powders condensed on the inner surface of the quartz tube and made the growth impossible to continue after a while. The deposit inside the quartz tube was found to be Pr$_6$O$_{11}$, which suggests that Pr$^{3+}$ is oxidizing to higher oxidation state. This oxidizing behavior suggests that air does not serve as a good atmosphere, thus urging for an inert atmosphere to be more suitable for the growth. From this growth attempt we could also assume that Pr$_2$Hf$_2$O$_7$  melts incongruently in air atmosphere. Therefore crystal growth using a self-flux, that can reduce the melting temperature, was thought be more suitable.

In our second attempt the growth was done by the traveling solvent floating zone technique, using a solvent with excess Pr$_2$O$_3$ (60 mol\% Pr$_2$O$_3$ and 40 mol\% Hf$_2$O$_4$). With excess Pr$_2$O$_3$ the eutectic melting point is lowered which allows growth at a relatively lower temperature. In order to compensate for the loss due to evaporation, we used a feed rod with 3\% excess Pr$_2$O$_3$ (3\% excess Pr$_6$O$_{11}$ was added to the starting composition of feed rod). The growth atmosphere was changed to argon in order to prevent Pr$_2$O$_3$ from oxidizing to higher oxidized state. Crystal growth was done at a rate of 1.2~mm/h. Under these conditions we observed significant reduction in evaporation during growth. Single crystal of Pr$_2$Hf$_2$O$_7$ could be grown by this solvent floating zone growth process, though the crystal was not of high quality. Further, the visual observation of the melting behavior of the feed-rod suggested that the compound possibly melts congruently in Ar atmosphere.

That the Pr$_2$Hf$_2$O$_7$ may be congruently melting in Ar atmosphere, prompted us to try floating zone growths (without traveling solvent) so that the growth rate can be increased, and large crystals can be grown in a relatively shorter time. Indeed, in Ar atmosphere we could successfully grow large single crystals by the floating zone method using the feed rods prepared from both stoichiometric and off-stoichiometric starting compositions. Subsequently, the crystals were grown under flowing argon atmosphere (2~L/h) by the floating zone technique. Crystal growths were done at a relatively higher speed of 5~mm/h.  Large single crystals of about 5~mm in diameter and 4--8~cm in lengths could be successfully grown by this process. 

We noticed that the appearance of the crystal depends on the stoichiometry of the starting material. We will be focussing here mainly on two crystals, one grown with a stoichiometric feed rod (crystal X1), and the other grown (optimally) with an offstoichiometric feed rod (crystal X2). The crystal X1 grown with stoichiometric feed rod (growth rate 5~mm/h, Ar amosphere) was bottle green in color [Fig.~\ref{fig:Laue}(a)]. The crystal X2 that was grown with feed rod having 3\% excess Pr$_2$O$_3$ (growth rate 5~mm/h, Ar amosphere) was fluorescent green in color [see Fig.~\ref{fig:Laue}(a)]. For crystal X2 the feed rod was prepared with the starting materials taken as: $1.03~\times$~Pr$_6$O$_{11}$ + $6~\times$~HfO$_2$ =  $3~\times($~Pr$_2$Hf$_2$O$_7$ + $0.03~\times$~Pr$_2$O$_3$) + $1.03~\times$~O$_2$. Growth conditions for crystal X2 was found to be optimum in terms of crystal quality and stoichiometry.

The as-grown crystal X1 and a piece of crystal X2 are shown in Fig.~\ref{fig:Laue}(a). Despite a clear change in the appearance of crystals, we could not notice any difference in the quality of crystals from X-ray Laue photographs. The representative Laue photographs for the three principal crystallographic directions  [100], [110] and [111] of optimally grown crystal X2 are shown in Fig.~\ref{fig:Laue}(b--d)). No difference was noticed in the laboratory based powder X-ray diffraction (data not shown) performed on the crushed single crystals X1 and X2.  We therefore performed synchrotron X-ray diffraction measurements to probe the microscopic details of crystals. The cross sections of the single crystals were examined by polarized optical microscopy and found to be free of grain boundaries or inclusions. The EDX composition analysis of single crystals revealed the stoichiometric ratio for Pr and Hf for both the crystals. However, due to the experimental limitations of EDX analyzer it was not possible to obtain any reliable estimate for light element oxygen.

\begin{figure}
\includegraphics[width=8.7cm, keepaspectratio]{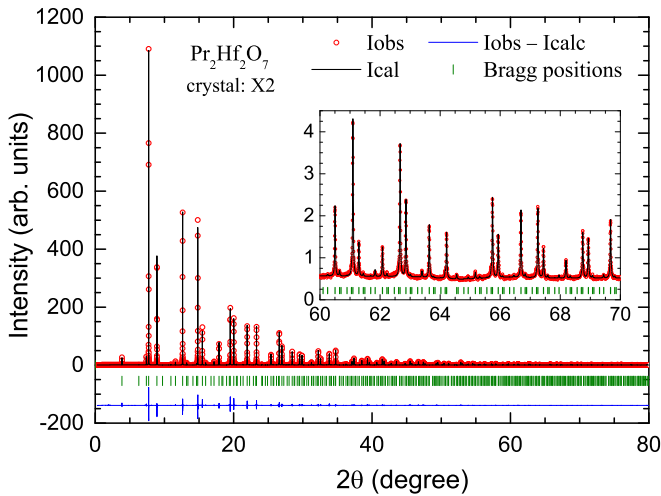}
\caption {Synchrotron powder x-ray diffraction pattern of crushed Pr$_2$Hf$_2$O$_7$ single crystal (\# X2) recorded at room temperature. The solid line through the experimental points is the Rietveld refinement profile calculated for the ${\rm Eu_2Zr_2O_7}$-type face-centered cubic (space group $Fd\bar{3}m$) pyrochlore structure. The short vertical bars mark the Bragg peaks positions and the lowermost curve represents the difference between the experimental and calculated intensities. Inset: An expanded scale view showing the details of the refinement in a small angle range at high $2\theta$.}
\label{fig:XRD}
\end{figure}

\begin{table}
\caption{\label{tab:XRD} Refined crystallographic parameters and agreement factors obtained from the structural Rietveld refinement of room temperature synchrotron x-ray diffraction measurements on powdered (optimally grown) crystal \# X2 of Pr$_2$Hf$_2$O$_7$. The Wyckoff positions (atomic coordinates) of Pr, Hf, O1 and O2 atoms in space group $Fd\bar{3}m$ are 16d (1/2,1/2,1/2), 16c (0,0,0), 48f ($x_{\rm O1}$,1/8,1/8) and 8b (3/8,3/8,3/8), respectively. The crystallographic parameters of polycrystalline Pr$_2$Hf$_2$O$_7$ are also listed for comparison \cite{Anand2016}.}
\begin{ruledtabular}
\begin{tabular}{lcc}
 & crystal (\# X2) & polycrystalline \\
 \hline
 \underline{Lattice parameters}\\
{\hspace{0.8cm} $a$ ({\AA})}            			&  10.6727(1) & 10.6728(1) \\	
{\hspace{0.8cm} $V_{\rm cell}$  ({\AA}$^{3}$)} 	&  1215.70(1)  &  1215.72(1)\\
\underline{Atomic coordinate}\\
\hspace{0.8cm} $x_{\rm O1}$ & 0.3353(3) & 0.3351(4)\\
\underline{Refinement quality} \\
\hspace{0.8cm} $R_{\rm p}$ (\%)  & 8.49 & \\
\hspace{0.8cm} $R_{\rm wp}$ (\%) & 10.1 &  \\
\hspace{0.8cm} $R_{\rm Bragg}$ (\%) & 4.57 & \\
\hline
\end{tabular}
\end{ruledtabular}
\end{table}

\section{\label{Structure} Crystallography}

Pr$_2$Hf$_2$O$_7$ forms in face-centered cubic  (space group $Fd\bar{3}m$) pyrochlore structure \cite{Karthik2012, Anand2016}. The crystal structure is illustrated in Fig.~\ref{fig:Struct}(a). In this pyrochlore structure, while Pr and Hf atoms occupy only one site: 16d (1/2,1/2,1/2) and 16c (0,0,0), respectively, the O atoms occupy two sites: 48f ($x_{\rm O1}$,1/8,1/8) and 8b (3/8,3/8,3/8), designated as  O1 and O2, respectively \cite{Subramanian1983}. The formula unit can therefore be viewed as Pr$_2$Hf$_2$O1$_6$O2. The Pr(16d) atoms are surrounded by 6 O1(48f) atoms and 2 O2(8b) atoms, and are thus eight-fold coordinated \cite{Subramanian1983}. The coordination unit is a scalenohedra PrO1$_6$O2$_2$ and the Pr atoms sit at the centre of a distorted cube formed by 6 O1 and 2 O2, the distance Pr--O2 being slightly smaller than Pr--O1.  The Hf(16c) atoms are surrounded by 6 O1(48f) atoms, and are thus six-fold cordinated \cite{Subramanian1983}. The coordination unit is a trigonal antiprism HfO1$_6$ and the Hf atoms sit at the centre of a distorted octahedra formed by 6 equally distant O1 atoms. Both Pr(16d) and Hf(16c) separately form three-dimensional networks of corner-sharing tetrahedra. The corner-sharing tetrahedra formed by Pr atoms is illustrated in Fig.~\ref{fig:Struct}(b).

The room temperature high-resolution synchrotron XRD pattern for powdered Pr$_2$Hf$_2$O$_7$ single crystal X2 is shown in Fig.~\ref{fig:XRD}. The structural Rietveld refinement profile using the software FullProf \cite{Rodriguez1993} is also shown in Fig.~\ref{fig:XRD}. The Rietveld refinement confirmed the ${\rm Eu_2Zr_2O_7}$-type face-centered cubic  (space group $Fd\bar{3}m$) pyrochlore structure. The refinement also reflects the single phase nature of the sample. The crystallographic and refinement quality parameters are listed in Table~\ref{tab:XRD}.  The lattice parameters agree very well with that of the polycrystalline Pr$_2$Hf$_2$O$_7$ \cite{Anand2016,Karthik2012}. No Pr/Hf site mixing and/or oxygen deficiency could be detected in our refinement, which if present, is expected to be less than 0.5\%.

\begin{figure} 
\includegraphics[width=\columnwidth, keepaspectratio]{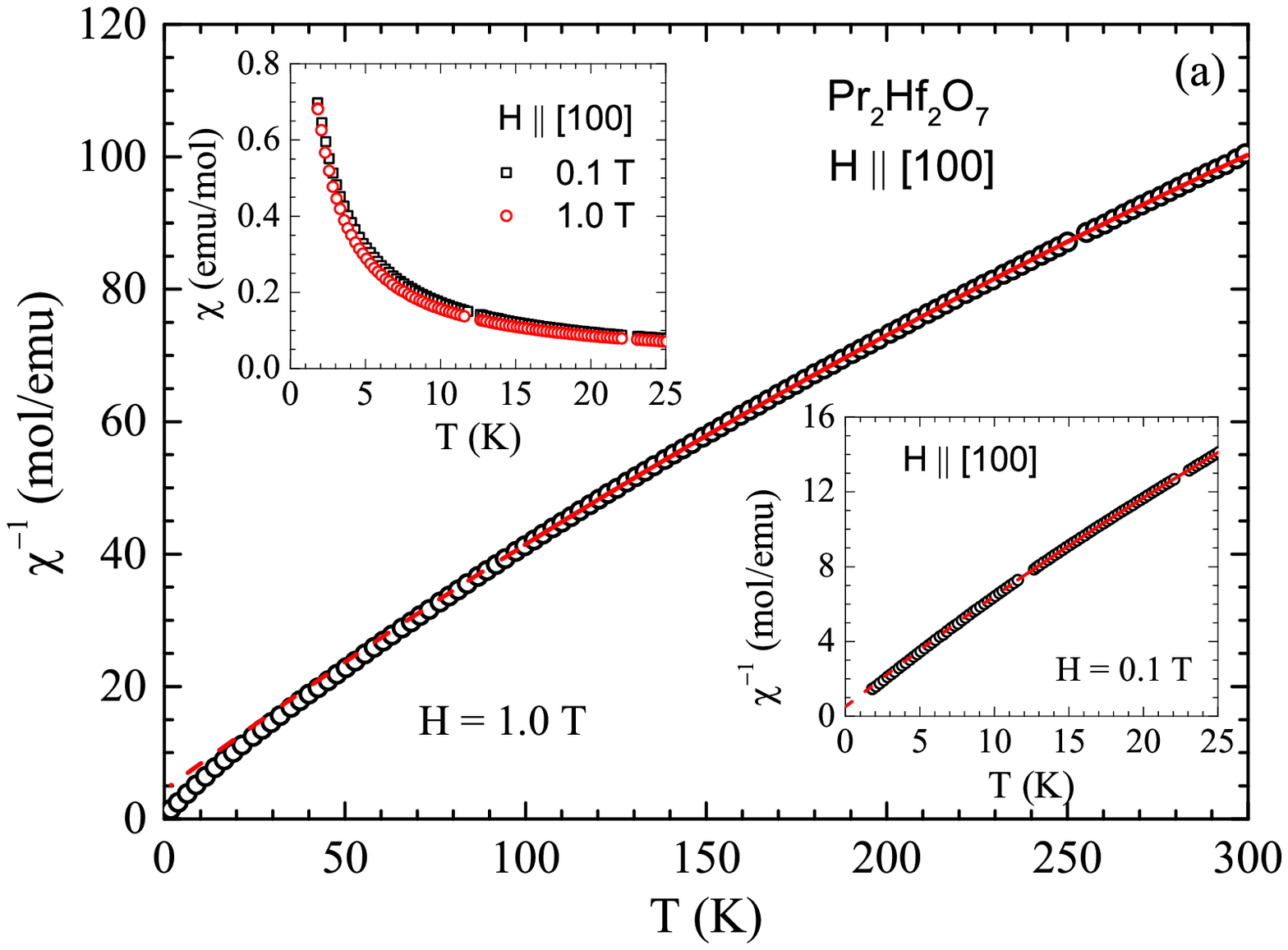}
\includegraphics[width=\columnwidth, keepaspectratio]{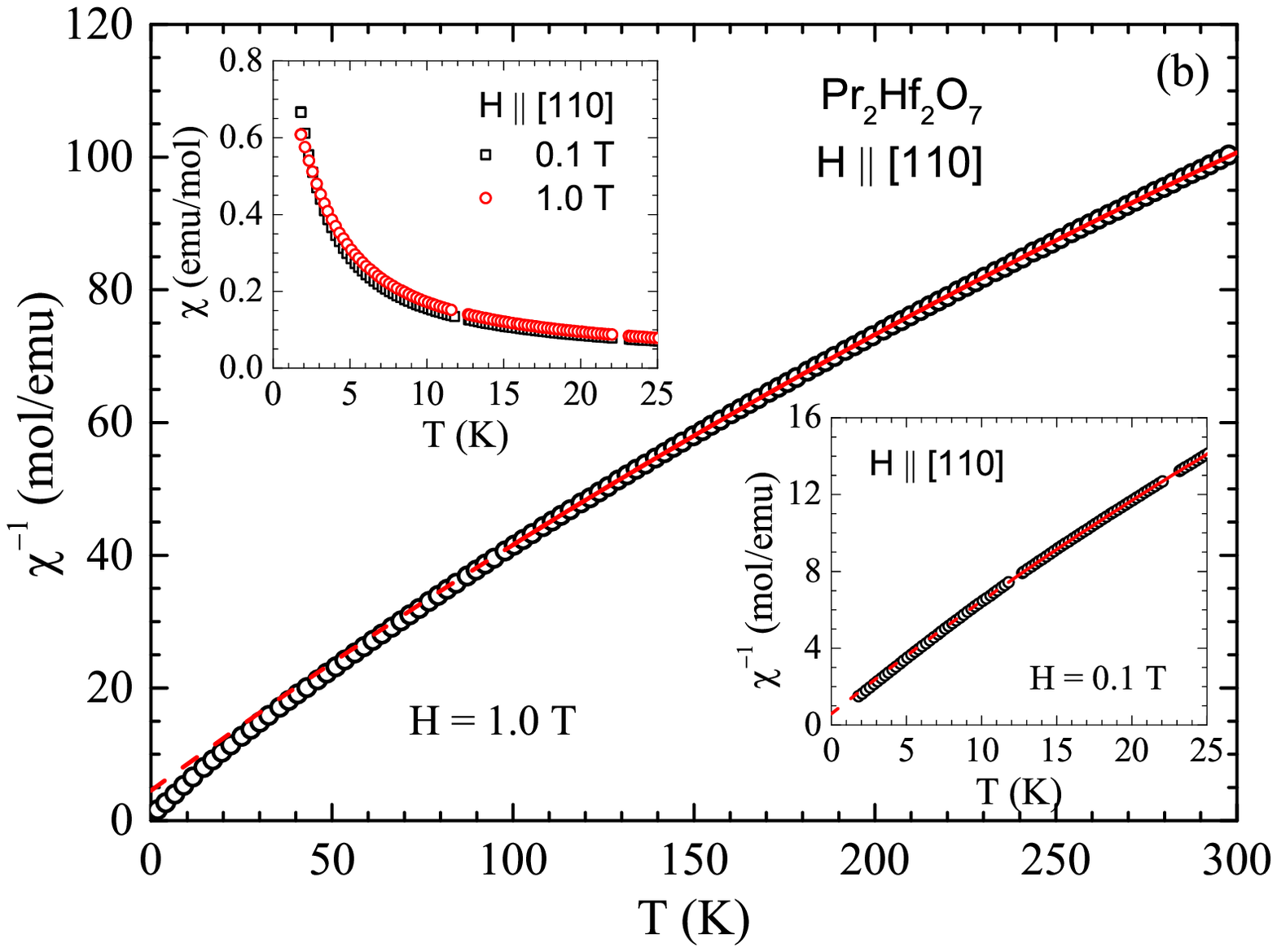}
\includegraphics[width=\columnwidth, keepaspectratio]{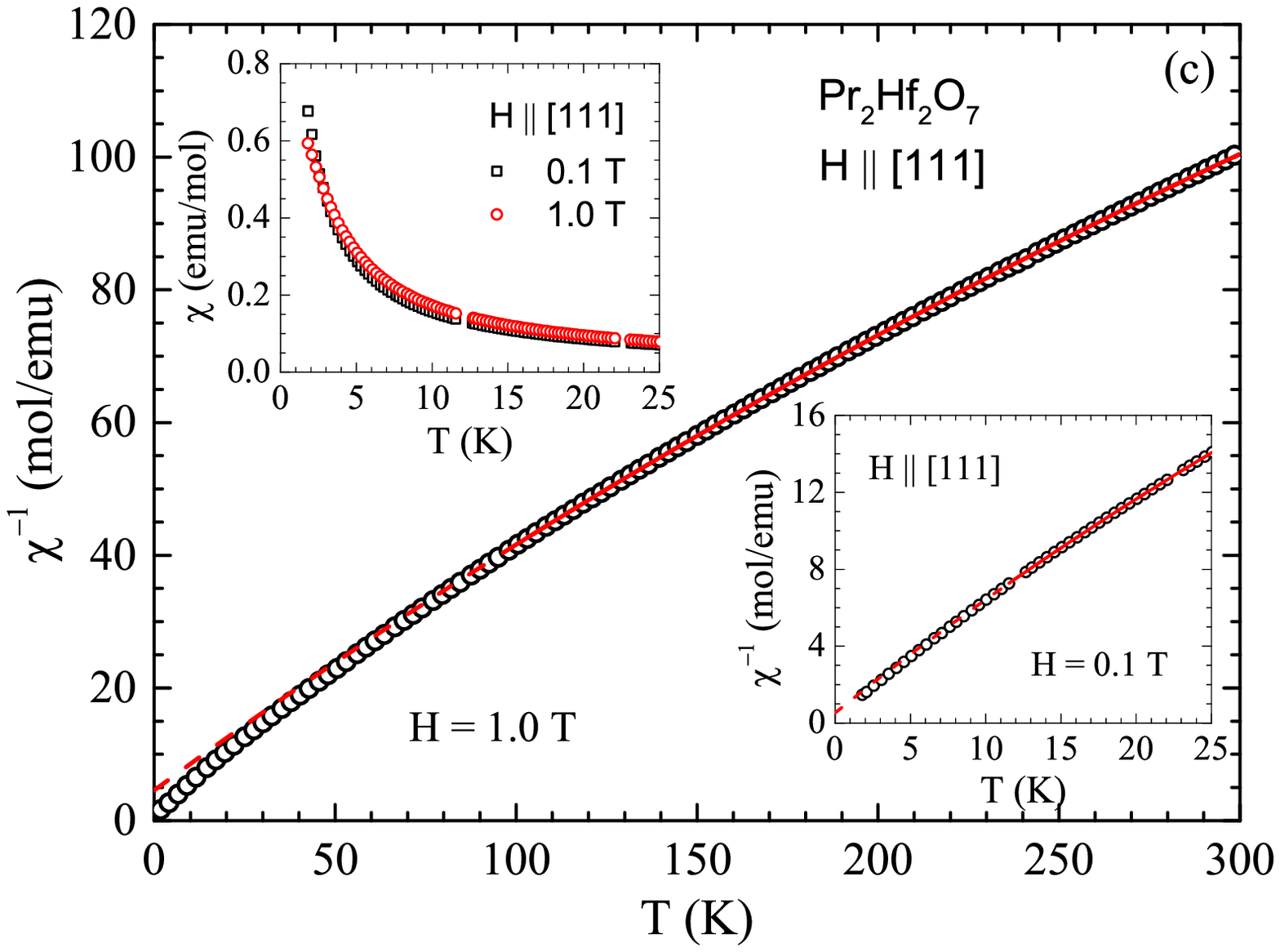}
\caption{Zero-field-cooled magnetic susceptibility $\chi$ of single crystal Pr$_2$Hf$_2$O$_7$ plotted as inverse susceptibility  $\chi^{-1}$ as a function of temperature $T$ for $1.8~{\rm K} \leq T \leq 300$~K for the three crystallographic directions [100], [110] and [111], measured in magnetic field $H= 1.0$~T\@. The solid red lines are the fits of the $\chi^{-1}(T)$ by the modified Curie-Weiss law in $100~{\rm K} \leq T \leq 300$~K and the dashed lines are the extrapolations. Upper insets: Low-$T$ $\chi(T)$ for $1.8~{\rm K} \le T \leq 25$~K measured in $H= 0.1$ and 1.0~T\@.  Lower insets: $\chi^{-1}(T)$ for $H=0.1$~T and $1.8~{\rm K} \le T \leq 25$~K\@ with the fits by the modified Curie-Weiss law in $12~{\rm K} \leq T \leq 25$~K.}
\label{fig:MT}
\end{figure}

The microscopic differences in the SXRD patterns of the single crystals X1 and X2 are compared in Fig.~\ref{fig:Struct}(d,e). For comparison the SXRD pattern of polycrystalline Pr$_2$Hf$_2$O$_7$ \cite{Anand2016} is also presented [Fig.~\ref{fig:Struct}(c)]. Peak splitting is clearly seen in the case of crystal X1 [see Fig.~\ref{fig:Struct}(d)] which is not there in the case of optimally grown crystal X2 [Fig.~\ref{fig:Struct}(e)] or in the case of the polycrystalline sample [Fig.~\ref{fig:Struct}(c)]. The peak splitting reflects a distribution of lattice parameters in crystal X1 due to the presence of sub-crystals of slightly different lattice parameters suggesting off-stoichiometry. On the other hand no such distribution of lattice parameter is present in the case of crystal X2 [Fig.~\ref{fig:Struct}(e)] for which the pattern is very similar to that of the polycrystalline sample [Fig.~\ref{fig:Struct}(c)]. Thus we see that the optimally grown crystal X2 is the best crystal.    

\begin{figure}
\includegraphics[width=8.7cm, keepaspectratio]{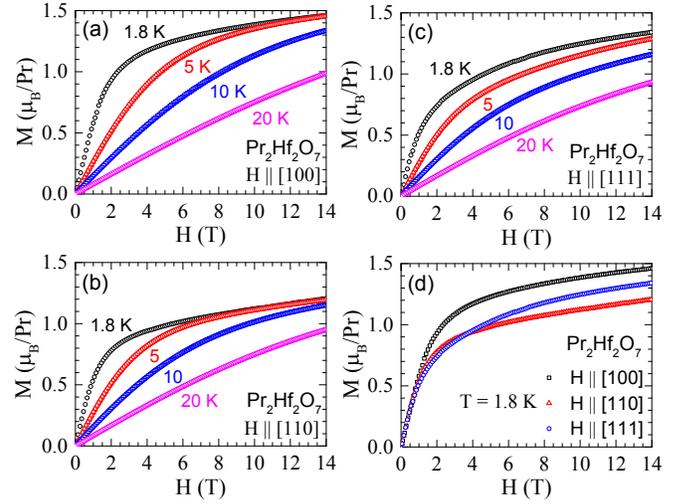}
\caption{Isothermal magnetization $M$ of single crystal Pr$_2$Hf$_2$O$_7$ as a function of applied magnetic field $H$ for $0 \leq H \leq 14$~T for the field (a) $H \parallel [100]$, (b) $H \parallel [110]$ and (c) $H \parallel [111]$ for the temperatures 1.8~K, 5~K, 10~K and 20~K. (d) A comparision of 1.8~K $M(H)$ data for three crystallographic directions [100], [110] and [111].}
\label{fig:MH}
\end{figure}

\section{\label{Sec:ChiMH} Magnetization and Heat Capacity}

The zero-field-cooled magnetic susceptibility $\chi(T)$ data of single crystal Pr$_2$Hf$_2$O$_7$ are shown in Fig.~\ref{fig:MT} for $H = 0.1$ and 1.0~T applied along the three crystallographic directions $H \parallel [100]$, $H \parallel [110]$ and $H \parallel [111]$. As can be seen from the upper insets of Fig.~\ref{fig:MT} the $\chi(T)$ for all $H \parallel [100]$, $H \parallel [110]$ and $H \parallel [111]$ are very similar in magnitude and do not show any anomaly at $T \geq 1.8$~K\@. The $\chi(T)$ follows modified Curie-Weiss behavior, $\chi(T) = \chi_0 + C/(T-\theta_{\rm p})$. We fitted the $\chi(T)$ data in $12~{\rm K} \leq T \leq 25$~K to get an estimate of $\theta_{\rm p}$ and $\mu_{\rm eff}$ for the Ising anisotropic ground state of Pr$_2$Hf$_2$O$_7$. The  fits of the $H = 0.1$~T $\chi^{-1}(T)$ data by the modified Curie-Weiss law over $12~{\rm K} \leq T \leq 25$~K are shown in the lower insets of Fig.~\ref{fig:MT}. The parameters obtained from the fits are listed in Table~\ref{tab:CW_chi}. An effective moment of $\mu_{\rm eff} \approx 2.51\, \mu_{\rm B}$/Pr is obtained from the values of $C$. The $\mu_{\rm eff}$ obtained is much lower than the free ion value of $3.58\, \mu_{\rm B}$/Pr for Pr$^{3+}$ ions, reflecting the presence of strong Ising anisotropy. 

\begin{table*}
\caption{\label{tab:CW_chi} Parameters obtained from the analysis of magnetic susceptibility of Pr$_2$Hf$_2$O$_7$ single crystal by modified Curie-Weiss law.} 
\begin{ruledtabular}
\begin{tabular}{ccccccc}
$H$ direction  & H  & Fit $T$ range      & $\chi_0$            &  $C$                   & $\theta_{\rm p}$ & $\mu_{\rm eff}$ \\
                     & (T)  & (K)      &  (emu/mol\,Pr)   & (emu\,K/mol\,Pr) &  (K) & $\mu_{\rm B}$/Pr \\
\hline
$[100]$    & 0.1  &$12~{\rm K} \leq T \leq 25$~K  & $4.9(1)   \times 10^{-3}$      & 0.79(1)      &   $-0.81(4)$  & 2.51   \\
$[110]$    & 0.1  &$12~{\rm K} \leq T \leq 25$~K & $4.8(1)   \times 10^{-3}$      & 0.79(1)      &   $-0.93(5)$   & 2.51   \\
$[111]$    & 0.1  &$12~{\rm K} \leq T \leq 25$~K &  $4.9(1)   \times 10^{-3}$     & 0.79(1)      &   $-0.86(5)$   & 2.51  \\
$[100]$    & 1.0  &$100~{\rm K} \leq T \leq 300$~K  & $1.07(4)   \times 10^{-3}$      &1.22(1)      &   $-10.7(2)$   & 3.12  \\
$[110]$    & 1.0  &$100~{\rm K} \leq T \leq 300$~K & $1.04(4)   \times 10^{-3}$      & 1.22(1)      &   $-11.2(3)$    & 3.12 \\
$[111]$    & 1.0  &$100~{\rm K} \leq T \leq 300$~K &  $1.06(3)   \times 10^{-3}$     & 1.22(1)      &   $-11.2(2)$   & 3.12  \\
\end{tabular}
\end{ruledtabular}
\end{table*}

We also fitted the high-$T$ $\chi(T)$ data with modified Curie-Weiss law yielding the values of  $\theta_{\rm p}$ and $\mu_{\rm eff}$ in the paramagnetic state. The  fits of $H = 1.0$~T $\chi^{-1}(T)$ data by the modified Curie-Weiss law in $100~{\rm K} \leq T \leq 300$~K are shown in Fig.~\ref{fig:MT}. The fit parameters are listed in Table~\ref{tab:CW_chi}. The $C$ value gives an effective moment of $\mu_{\rm eff} \approx 3.12\, \mu_{\rm B}$/Pr which is again smaller than the expected value of $3.58\, \mu_{\rm B}$/Pr for free Pr$^{3+}$ ions. The smaller value of $\mu_{\rm eff}$ could be the result of strong crystal field effect. As the first excited CEF state in Pr$_2$Hf$_2$O$_7$ lies only at 9.1~meV \cite{Anand2016}, both the ground state doublet and the excited singlet are thermally occupied in this temperature range. Thus the estimate of $\mu_{\rm eff}$ may be affected by the thermal population of this CEF level. 

\begin{figure}
\includegraphics[width=8.7cm, keepaspectratio]{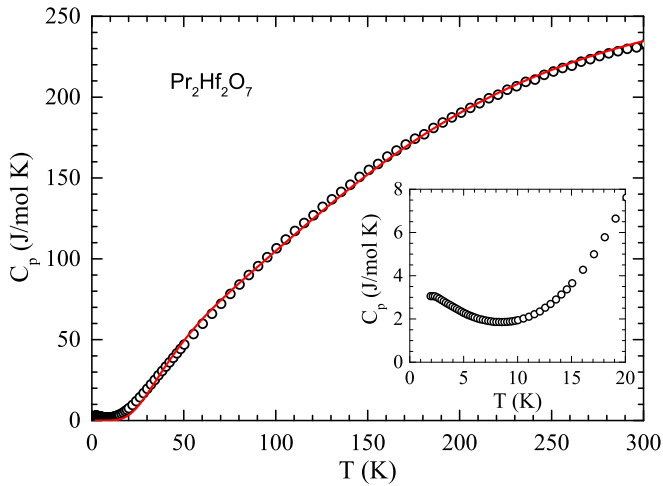}
\caption{\label{fig:HC} Heat capacity $C_{\rm p}$ of single crystal Pr$_2$Hf$_2$O$_7$ as a function of temperature $T$ for 1.8~K~$\leq T \leq$~300~K measured in zero field. The solid curve is the fit of $C_{\rm p}(T)$ by Debye+Einstein models of lattice heat capacity plus crystal field contribution as deduced from inelastic neutron scattering
(INS) data \cite{Anand2016}. Inset: Expanded view of \mbox{low-$T$} $C_{\rm p}(T)$ data over $1.8~{\rm K} \leq T \leq 20$~K\@.} 
\end{figure}

The isothermal magnetization $M(H)$ data of single crystal Pr$_2$Hf$_2$O$_7$ are shown in Fig.~\ref{fig:MH} for $H$ along the three crystallographic directions, i.e. $H \parallel [100]$, $H \parallel [110]$ and $H \parallel [111]$ at $T= 1.8$~K, 5~K, 10~K and 20~K. It is seen that the $M(H)$  isotherms for all three directions of $H$ become nonlinear at lower temperatures reflecting the saturation tendency of magnetization in a mean-field way. A comparison of the $M(H)$ isotherms for $H \parallel [100]$, $H \parallel [110]$ and $H \parallel [111]$ at 1.8~K is shown in Fig.~\ref{fig:MH}(d). The $M(H)$ data show clear anisotropy with $M_{[100]} > M_{[111]} > M_{[110]}$ as expected for Ising anisotropic behavior in 227 pyrochlores. The $M(H)$ data follow saturation values of $g_{\rm eff} J_{\rm eff} \mu_{\rm B}(1/\sqrt 3)$ for [100] (two-in/two-out), $g_{\rm eff} J_{\rm eff}\mu_{\rm B}(\sqrt{2/3} \times2)/4$ for [110] (one-in/one-out, two free) and $g_{\rm eff} J_{\rm eff}\mu_{\rm B}(1+1/3\times 3)/4$ for [111] (three-in/one-out) as expected for a system with $\langle 111 \rangle$ Ising anisotropy \cite{Fukazawa2002}. As seen from the Fig.~\ref{fig:MH}(d), at $H=14$~T the values of $M$ are $M_{[100]} = 1.46\,\mu_{\rm B}$/Pr, $M_{[110]} = 1.21 \,\mu_{\rm B}$/Pr and $M_{[111]} = 1.34 \,\mu_{\rm B}$/Pr which are consistent with the estimated Ising anisotropic values $M_{[100]} = 1.47\,\mu_{\rm B}$/Pr, $M_{[110]} = 1.04 \,\mu_{\rm B}$/Pr and $M_{[111]} = 1.27 \,\mu_{\rm B}$/Pr taking $g_{\rm eff} J_{\rm eff} \mu_{\rm B} = 2.54~\, \mu_{\rm B} $, the effective moment. The observed higher value of $M_{[110]}$ possibly indicates the presence of non-Ising contributions. Further we also notice that in contrast to the expected saturation behavior for Ising anisotropic spin-ice materials, the $M$ for all three directions keeps on increasing up to the measured field of 14~T and this weak field dependence is again a signature for the presence of non-Ising term in the Hamiltonian.  

The heat capacity $C_{\rm p}(T)$ data of Pr$_2$Hf$_2$O$_7$ single crystal for 1.8~K~$\leq T \leq$~300~K are shown in Fig.~\ref{fig:HC}. As can be seen from the inset of Fig.~\ref{fig:HC} the low-$T$  $C_{\rm p}(T)$ data show an upturn at $T \leq 10$~K which is a residual of peak near 1.8~K \cite{Sibille2016} associated with the slow spin dynamics. The high-$T$ $C_{\rm p}(T)$ data are well represented by  the Debye and Einstein models of lattice heat capacity.  The fit of $C_{\rm p}(T)$ data by Debye and Einstein models of heat capacity as detailed in Ref. \citenum{Anand2015a} plus crystal field contribution deduced from the inelastic neutron scattering (INS) measurement on  Pr$_2$Hf$_2$O$_7$ \cite{Anand2016} is shown by the solid red curve in Fig.~\ref{fig:HC}. The fit gives $\Theta_{\rm D} = 788(6)$~K and Einstein temperature $\Theta_{\rm E} = 160(2)$~K with 66\% weight to Debye term and 34\% to Einstein term. 

We note that the magnetic susceptibility, isothermal magnetization and heat capacity measurements on crystal X1 and X2 did not reveal any noticeable difference within the accuracy of instruments used. The bulk properties of optimized crystal X2 are consistent with those of polycrystalline sample \cite{Anand2016}. Our magnetic susceptibility and isothermal magnetization data are also in good agreement with the recently reported data of single crystal Pr$_2$Hf$_2$O$_7$ \cite{Hatnean2017}.

\section{\label{Conclusion} Summary and Conclusions}

We have successfully grown the large single crystals of candidate quantum spin-ice Pr$_2$Hf$_2$O$_7$ using optical floating-zone method and optimized the growth conditions. Further, the microscopic structural differences of crystals grown under different conditions were probed by the high-resolution synchrotron x-ray diffraction which revealed that the crystal grown under the optimized condition is free from sub-lattice formation or site mixing. The best crystal was obtained with starting material having 3\% excess Pr$_6$O$_{11}$, grown at a growth rate of 5~mm/hour under flowing argon atmosphere ($\sim 2$~L/h). As the presence of distribution of lattice parameters can affect the quantum spin ice physics, the optimally grown crystals should provide a scope for further study of the intrinsic properties of quantum spin-ice candidate Pr$_2$Hf$_2$O$_7$. The optimally grown crystal was investigated by the magnetic susceptibility, isothermal magnetization and heat capacity measurements. The $\chi(T)$ and $M(H)$ data confirm the Ising anisotropy and presence of weak non-Ising contribution as previously observed for polycrystalline Pr$_2$Hf$_2$O$_7$. 

\section*{Acknowledgements}

We acknowledge Helmholtz Gemeinschaft for funding via the Helmholtz Virtual Institute (Project No. VH-VI-521).

\end{document}